# Universal Patterns in the Blockchain: Analysis of EOAs and Smart Contracts in ERC20 Token Networks

Kundan Mukhia [1], SR Luwang [1] Md. Nurujjaman[1‡] Tanujit Chakraborty [2,3], Suman Saha[4*], Chittaranjan Hens[5]

**1** Department of Physics, National Institute of Technology, 737139, Sikkim, India
**2** SAFIR, Sorbonne University, Abu Dhabi, UAE
**3** Sorbonne Center for Artificial Intelligence, Sorbonne University, Paris, France
**4** School of Electronics Engineering, Vellore Institute of Technology, Chennai, Tamil Nadu 600127, India
**5** Centre for Computational Natural Sciences and Bioinformatics, International Institute of Information Technology, Hyderabad 500032, India.

kundanmukhia07@gmail.com  salamrabindrajit@gmail.com
‡md.nurujjaman@nitsikkim.ac.in  tanujit.chakraborty@sorbonne.ae
*suman.saha@vit.ac.in  chittaranjanhens@gmail.com

## Abstract

Scaling laws offer a powerful lens to understand complex transactional behaviors in decentralized systems. This study reveals distinctive statistical signatures in the transactional dynamics of ERC20 tokens on the Ethereum blockchain by examining over 44 million token transfers between July 2017 and March 2018 (9-month period). Transactions are categorized into four types: EOA–EOA, EOA–SC, SC-EOA, and SC-SC based on whether the interacting addresses are Externally Owned Accounts (EOAs) or Smart Contracts (SCs), and analyzed across three equal periods (each of 3 months). To identify universal statistical patterns, we investigate the presence of two canonical scaling laws: power law distributions and temporal Taylor's law (TL). EOA-driven transactions exhibit consistent statistical behavior, including a near-linear relationship between trade volume and unique partners with stable power law exponents ($\gamma \approx 2.3$), and adherence to TL with scaling coefficients ($\beta \approx 2.3$). In contrast, interactions involving SCs, especially SC-SC, exhibit sublinear scaling, unstable power-law exponents, and significantly fluctuating Taylor coefficients (variation in $\beta$ to be $\Delta\beta = 0.51$). Moreover, SC-driven activity displays heavier-tailed distributions ($\gamma < 2$), indicating bursty and algorithm-driven activity. These findings reveal the characteristic differences between human-controlled and automated transaction behaviors in blockchain ecosystems. By uncovering universal scaling behaviors through the integration of complex systems theory and blockchain data analytics, this work provides a principled framework for understanding the underlying mechanisms of decentralized financial systems.

## Introduction

Blockchain technology has revolutionized the structure of financial systems by enabling decentralized, trustless transactions through distributed ledgers. The rapid growth in cryptocurrencies has significantly reshaped the global financial landscape. Since the launch of Bitcoin in 2009 [1], the number of cryptocurrencies has expanded dramatically [2, 3, 4, 5]. As of April 2025, there are approximately 25,000 active cryptocurrencies, including both coins and tokens across various blockchain platforms. Together, they hold a total market capitalization of around USD 2.96 trillion [6]. Cryptocurrencies have gained widespread attention due to their decentralized nature, lower transaction costs compared to traditional fiat currencies, and the transparency they offer [7, 8, 9]. At the core of cryptocurrencies lies blockchain technology, a distributed append-only ledger that securely records transactions [9, 10]. Blockchains often operate with native cryptocurrencies, which can be traded for traditional money through exchanges like Binance [11]. The ability to freely exchange cryptocurrencies has significantly fueled their market growth. It has also enabled new fundraising methods, such as Initial Coin Offerings (ICOs), where blockchain-based digital tokens are sold directly to participants [12, 13].

Among various blockchain platforms, Ethereum has emerged as a dominant platform, supporting programmable smart contracts and the proliferation of ERC20 tokens that underpin a wide range of decentralized financial applications [12]. To facilitate token development and exchange, the Ethereum community introduced the ERC20 token standard in November 2015 [14]. Although it is not mandatory, the ERC20 standard has been adopted widely, with the majority of Ethereum-based tokens being ERC20 compatible [13]. One of the unique features of Ethereum is its dual-account model, consisting of Externally Owned Accounts (EOAs), which represent individual users [12], and smart contracts (SCs), which are autonomous programs executing predefined logic [15, 16]. This structure creates two different types of interactions on the Ethereum network: *Human-driven (EOA-based) and Algorithm-driven (SC-based) transactions*. These distinct modes of operation give rise to heterogeneous behavioral patterns across the network. As the volume and complexity of blockchain activity continue to grow, understanding the underlying transactional dynamics, particularly the differences between human and automated agents, becomes increasingly essential for ensuring transparency.

Several studies have explored ERC20 token transactions from various angles. For instance, Victor et al. [8] analyzed the wash trading activity on decentralized exchanges like IDEX and EtherDelta and found that over 30% of the tokens traded were involved in wash trading, indicating manipulative behavior. Pradeep et al. [17] examined the behavior of ERC20 token traders during the 2018 crypto crash and the COVID-19 pandemic, revealing that trader interactions became more diverse in response to these market shocks, reflecting increased adoption of ERC20 tokens. Chen et al. [12] investigated the broader ERC20 ecosystem, raising important concerns about market manipulation and token price dynamics. In addition, Victor et al. [13] identified hub-and-spoke patterns in token transaction networks and observed that many tokens were primarily sold rather than circulated, highlighting imbalances in token flow. While these studies provide valuable insights into the ERC20 ecosystem, they generally treat token transactions as a unified whole, without explicitly differentiating between human-driven and algorithm-driven interactions. However, such a distinction is crucial for understanding how EOAs SCs operate differently, which could unveil deeper structural and behavioral patterns in token transactions. Despite the dual-account model of Ethereum being central to its design, there remains a significant research gap in developing robust frameworks that systematically distinguish between human users and algorithmic agents to better interpret the heterogeneous dynamics of ERC20 token activity.

Earlier blockchain studies have identified power-law patterns in transaction networks, reflecting scalable and heterogeneous interactions [18, 19, 20]. Somin et al. [19] demonstrated that transaction volumes in ERC20 networks follow power-law distributions, akin to those observed in social [21] and biological networks [22]. Wu et al. [20] fur-ther showed that the market capitalizations of coins and tokens also exhibit power-law behavior. While power laws describe scalable transaction patterns, another canonical scaling law Taylor's law (TL) - measures fluctuation stability. Originally introduced to describe the relationship between the mean and variance of population sizes in ecological systems [23], TL has since been applied across a range of domains. In particular, it has gained prominence in economic and financial contexts as a tool for quantifying volatility and dynamic fluctuations in market activity [24, 25]. Despite the relevance of both scaling laws, no prior study has combined these frameworks to dissect the behavioral signatures of human-driven (EOA) and algorithmic (SC) interactions in blockchain ecosystems. Our study bridges this gap by integrating power-law analysis with TL to investigate ERC20 token transactions. While power-law exponents help reveal whether transactional dynamics are dominated by human or automated agents, Taylor's law provides insight into the stability of these interactions over time. Together, these methods offer a unified framework for distinguishing behavioral modes in decentralized systems, enabling a more granular understanding of ERC20 transaction types.

To address this research gap, we classify ERC20 transactions into four distinct categories based on the nature of interaction: human-to-human (EOA–EOA), human-to-contract (EOA–SC), contract-to-human (SC–EOA), and contract-to-contract (SC-SC). We investigate whether consistent statistical regularitiessuch as power-law scaling and fluctuation relationshipsemerge across these interaction types. Specifically, we analyze the scaling relationship between trade volume and partner diversity, fit power-law models to transaction distributions, and apply TL to assess the stability and persistence of scaling behavior over time. Our findings uncover clear statistical contrasts between human-driven and algorithm-driven activities in ERC20 transactions, offering deeper insights into the operational dynamics of decentralized financial systems. Beyond categorizing transactional roles, this study emphasizes how decentralized finance (DeFi) ecosystems are shaped by the interplay of human decisions and autonomous algorithmic executions. As DeFi platforms increasingly rely on smart contracts for lending, trading, and asset management, understanding these interaction types  purely human, purely algorithmic, and hybrid  is crucial for interpreting their impact on market stability, liquidity patterns, and the emergence of complex behaviors in blockchain-based economies.

The remainder of the paper is organized as follows: Section 1 details the materials and methods used in our study. Section 1.1 presents and discusses the results of our empirical analysis. Finally, Section 1.1 offers concluding remarks and outlines directions for future research.

# 1 Materials and Methods

## 1.1 ERC20 Transaction Dataset

We utilized nine months of historical ERC20 token transaction data from the Ethereum blockchain [26, 27]. The dataset includes a total of 44,858,196 token transfer records spanning the period from July 2017 to March 2018. Each record represents a token transfer between two addresses, identified by the columns from (sender) and to (receiver). Every transaction also contains a timestamp recorded in UTC format [28]. For instance, a timestamp such as 1512086400 corresponds to the date 2017-12-01, using the Unix epoch base time of 1970-01-01 00:00:00 UTC. These timestamps were converted into a human-readable format (YYYY-MM-DD) for temporal segmentation and trend analysis.

There are two types of accounts available in ERC20: one is EOAs, which are controlled by individuals using public-private key pairs, i.e., human users [12]. The other is SCs, which are governed by executable code stored within the account itself [15, 16]. We categorize the ERC20 transaction data based on whether the sender and receiver are EOAs or SCs using the columns fromIsContract and toIsContract, as shown in Table 1. Transactions between two EOAs are indicated by fromIsContract = 0 and toIsContract = 0. The four possible transaction categories based on sender and receiver types are categorically defined in Table 1. Further, we divided the total nine-month ERC20 transaction data into three equal periods: Period 1, Period 2, and Period 3, each spanning three months. The number of transactions for each period and the total number of transactions are presented in Table 1. This division helps us examine the temporal consistency and trends across the four transaction categories: EOA–EOA, EOA–SC, SC–EOA, and SC–SC. This temporal segmentation also enables a comparative analysis of evolving patterns across human-driven and algorithmic interactions.

Next, a scaling law has been verified between trade volume and partner diversity using a least squares linear regression fit on the log-log scatterplot, deriving the scaling exponent. Additionally, we perform power-law analysis on the trading activities using maximum likelihood estimation (MLE), verified using the Kolmogorov–Smirnov (KS) tests. Before applying Taylor's law (TL), we conduct the Kwiatkowski–Phillips–Schmidt–Shin (KPSS) test to ensure that the trading time series data is stationary. Further, we use TL to study how the variability of trading activity scales with the mean activity by fitting a variance-to-mean power-law distribution. Detailed descriptions and definitions of each of these metrics are discussed in the following sections.

**Table 1. Classification of ERC20 token transactions based on the sender and receiver account types with the transactional volume shown for the full dataset (Total Transaction), divided in three equal time periods.**

| fromisContract | toisContract | Classification | Period 1 | Period 2 | Period 3 | Total Transaction |
|---|---|---|---|---|---|---|
| 0 | 0 | EOA to EOA | 4028192 | 9730868 | 16685273 | 30444333 |
| 0 | 1 | EOA to SC | 629057 | 940365 | 1296837 | 2866259 |
| 1 | 0 | SC to EOA | 1673085 | 3856826 | 4830180 | 10360091 |
| 1 | 1 | SC to SC | 153399 | 446789 | 587325 | 1187513 |

The columns 'fromIsContract' and 'toIsContract' indicate whether the sender and receiver are externally owned accounts (EOAs, denoted by 0) or smart contracts ( SCs denoted by 1). This categorization forms the basis for analyzing transactional behavior across human-controlled and automated accounts.

## Scaling Law

To understand how individual trading activity relates to ERC20 transaction interactions, we examine the relationship between the number of unique trade partners and the total number of transactions for each trader. For every sender $i$, we calculate two quantities: the total number of transactions, denoted as $V_i = \sum_j v_{ij}$, and the number of distinct receivers, denoted as $N_i = |\{j\}|$, where $v_{ij}$ represents the number of times trader $i$ sent tokens to partner $j$. We investigate whether these quantities exhibit a power-law relationship, a pattern frequently observed across various complex systems [29, 30, 31, 32]. This relationship is typically expressed in the form:

$$V \sim N^\alpha, \tag{1}$$

where $\alpha$ is the scaling exponent that quantifies how transaction volume grows with the number of distinct partners. When $\alpha = 1$, the relationship is linear, suggesting that traders distribute their activity evenly across partners. If $\alpha > 1$, a superlinear trend is observed, indicating that more connected traders engage in disproportionately higher trade volumes. Conversely, a sublinear exponent ($\alpha < 1$) implies that trading activity increases more slowly compared to partner growth. We estimate the scaling exponent $\alpha$ by performing a linear fit on the log-transformed data, using the equation:

$$\log V = \alpha \log N + C, \qquad (2)$$

where $C$ is a constant. This method helps us uncover how trading behavior changes against partner diversity across different transaction types. This may offer greater insights into whether transactional growth follows predictable scaling regimes that can inform network design and congestion management in blockchain systems.

## Power Law Distribution

Power-law distributions are characterized by heavy tails, where the probability of large events decays as a power of the event size [33]. This property makes them well-suited for modeling a wide range of real-world phenomena [34]. Their scale-invariant nature implies a form of universality, suggesting that similar statistical patterns can emerge across different systems, regardless of their underlying mechanisms [35, 36, 37, 38]. The probability density function of the power-law distribution [33] is mathematically expressed as:

$$p(x) = \frac{x^{-\gamma}}{\zeta(\gamma, x_{\min})}, \quad x \geq x_{\min}. \qquad (3)$$

Here, $x$ denotes the observed value of the variable in this study, which represents the total number of trades received or sent by a trader. The parameter $\gamma$ is the scaling exponent that governs the rate at which probabilities decay as $x$ increases. The term $x_{\min}$ specifies the minimum threshold above which the power-law behavior is assumed to hold. The normalization factor $\zeta(\gamma, x_{\min})$ is the Hurwitz zeta function, ensuring that the distribution sums to one. The value of $\gamma$ plays a crucial role in determining the heaviness of the distribution's tail, with smaller values indicating more extreme deviations and heavier tails. In blockchain transactions, identifying power-law behavior helps us characterize whether universal mechanisms govern trading activity and whether human-driven and automated interactions follow distinct statistical regimes. The presence of power-law behavior in blockchain transactions indicates scale-invariant, self-organizing dynamics where a few highly active addresses dominate network activity.

## Maximum Likelihood Estimation (MLE) for Power Law Fit

To estimate the power-law parameters, we employed the maximum likelihood estimation (MLE) method [39, 40]. The parameter values are estimated by maximizing the likelihood function, representing the probability that the assumed power-law model generated the observed data. In essence, MLE identifies the parameter values that make the observed data most probable under the model.

The MLE provides an unbiased approach to parameter estimation, avoiding potential distortions that might occur when using log-log plots to fit a power-law. The MLE offers an accurate estimation of the parameters [41] and is complemented by the Kolmogorov-Smirnov (KS) test to evaluate the goodness-of-fit. All these combinations make the MLE a robust tool for estimating parameters for a power-law fit in complex systems.

## Determining Threshold and Model Evaluation

We determined the threshold value ($x_{min}$) for the power-law behavior using an algorithm that minimizes the Kolmogorov–Smirnov (KS) distance between empirical and theoretical cumulative distribution functions (CDFs) [42, 43, 44]. The KS distance is a widely adopted metric in power-law analysis, as it quantifies the maximum deviation between the observed data and the fitted model. This makes it particularly effective for selecting an appropriate $x_{\min}$. Formally, the KS distance is defined as:

$$D = \max_{x \geq x_{\min}} |S(x) - P(x)|,$$

where $S(x)$ is the empirical CDF of the data, and $P(x)$ is the CDF of the fitted power-law model. The KS statistic $D$ serves as a measure of goodness-of-fit: values below 0.05 indicate a close fit to the power-law distribution, while values above 0.1 suggest substantial deviation from power-law behavior.

**Log-Likelihood Ratio (LLR) Test**

The scaling exponent $\gamma$ in Eq. 3 is estimated using the MLE method, identifying the parameter value that makes the observed data most likely under the assumed power-law model. The uncertainty in this estimate is captured by the standard error $\sigma_\gamma$, where larger values of $\sigma_\gamma$ indicate higher estimation variability, typically resulting from limited data or weak adherence to the power-law form. To compare the power-law model against alternative distributions, such as an exponential distribution, we employed the log-likelihood ratio (LLR) test [45, 46], defined as:

$$\mathcal{R} = \sum_{i=1}^{n} \ln \left( \frac{p_{\text{power-law}}(x_i)}{p_{\text{exponential}}(x_i)} \right). \tag{4}$$

A positive value of $\mathcal{R}$ indicates that the power-law model is favored, while a negative value supports the exponential alternative. To assess the statistical significance of this preference, we compute a $p$-value, where $p < 0.05$ implies a statistically significant result. Collectively, these methods provide a robust and reproducible analytical framework for validating power-law behavior through parameter estimation, goodness-of-fit assessment, and model comparison.

## Stationarity Test

Time series analysis often begins with determining whether the data is stationary, a key property that influences model selection and interpretation. One of the most common approaches to assess stationarity is unit root testing. Several statistical tests have been developed to detect the presence of unit roots in time series data [47, 48, 49, 50, 51, 52].

We applied the Kwiatkowski-Phillips-Schmidt-Shin (KPSS) test, a widely used method for testing stationarity in time series analysis, particularly in econometrics [53, 54, 55, 56]. Unlike traditional unit root tests that assume non-stationarity as the null hypothesis, the KPSS test takes stationarity as the null hypothesis. The test is specified as:

$$Y_t = X_t + \beta t + \varepsilon_t, \quad t = 1, \ldots, T \tag{5}$$

$$X_t = X_{t-1} + u_t \tag{6}$$

where $\varepsilon_t$ is a stationary error term, $\beta t$ captures a deterministic trend, and $u_t$ is an independent and identically distributed (i.i.d.) noise term with zero mean and constant variance, i.e., $u_t \sim \text{iid}(0, \sigma^2)$. To perform the KPSS test, the Lagrange Multiplier (LM) statistic is computed as:

$$\hat{L}_M = \frac{1}{T^2} \sum_{t=1}^{T} S_t^2 / s^2(l), \tag{7}$$

where $S_t$ is the cumulative sum of residuals from the regression. The long-run variance $s^2(l)$ is estimated as:

$$s^2(l) = \frac{1}{T}\sum_{t=1}^{T}\hat{\varepsilon}_t^2 + \frac{2}{T}\sum_{s=1}^{l} w(s,l) \sum_{t=s+1}^{T}\hat{\varepsilon}_t\hat{\varepsilon}_{t-s}, \tag{8}$$

where $w(s,l) = 1 - \left(\frac{s}{l+1}\right)$ are the Bartlett weights used to smooth the autocovariance terms. The lag truncation parameter $l$ is selected using the Newey-West automatic bandwidth method [57]. To interpret the KPSS results, the LM statistic is compared against critical values, and a corresponding p-value is computed. If the p-value is greater than 0.05, we fail to reject the null hypothesis of stationarity, suggesting that the time series is stationary. Conversely, a p-value less than 0.05 leads to the rejection of the null hypothesis, indicating that the series is non-stationary.

**Temporal Taylor's Law**

Taylor's law (TL) or *fluctuation scaling* is a widely observed empirical pattern that relates the variance to the mean of groups of measurements or other non-negative quantities via a power law [23, 58, 59]. Originally discovered in ecological studies, it has since been observed across diverse disciplines including physics, finance, network science, and social systems [24, 59]. In this study, we have used temporal TL to examine how the variance of transaction activity relates to the mean activity over time for each account. To analyze this, we bin the transaction data into 1-hour time windows, counting the number of transactions each account makes in each period. For each account, we calculate the mean and variance of these hourly transaction counts. In ecology, Taylor's power law [23] states that the variance of a population is related to its mean through a power-law relationship, where the variance increases as a power of the mean, expressed as:

$$\sigma^2 = a\mu^b, \tag{9}$$

where $\mu$ is the mean of the measured quantity, $\sigma^2$ is the variance, $a$ is a positive constant, and $b$ is the TL exponent. The slope $b$ characterizes the degree of heterogeneity in the spatial or temporal patterns of the system [23, 60, 61]. Depending on the value of $b$, the system behavior can be interpreted as follows:

- $b = 1$: indicates Poisson-like (random) fluctuations. This suggests relatively stable behavior typical of random, uncoordinated human or contract interactions.

- $b > 1$: implies aggregation or clustering behavior. This suggests that it is possibly due to speculative activity, coordinated smart contracts, or market events triggering heavy trading.

- $b < 1$: reflects ordered or more uniform distributions. This may be observed where volatility grows more slowly than the mean, which can be seen in automated protocols or liquidity bots with bounded activity.

After a logarithm transformation, Eq. 9 can be written as:

$$\log(\sigma^2) = \log(a) + b\log(\mu), \tag{10}$$

This linear relationship allows the exponent $b$ to be easily estimated using methods, e.g., least squares linear regression fit. The consistent manifestation of this linear relationship across numerous systems has prompted researchers to propose it as a potential universal law in complex systems [24]. Temporal TL analysis enables us to distinguish human-driven interactions, which may be event-driven, from smart contract interactions, which may exhibit high-frequency and volatile behavior depending on protocol logic.

# Results

This section presents the results of scaling behaviors observed in ERC-20 token transactions by distinguishing interactions between Externally Owned Accounts (EOAs) and Smart Contracts (SCs). We classify all transactions into four categories: EOA–EOA, EOA–SC, SC–EOA, and SC–SC based on the roles of the interacting accounts. This categorization enables a detailed analysis of human-driven versus contract-driven dynamics. We investigate whether consistent statistical patterns, such as power-law distributions and Taylor's law (TL), emerge across these categories.

## Scaling Relationship between Trade Volume and Partner Diversity

Figures 1(a–d) show scatter plots in log-log scale of the number of unique trade partners versus the number of trades executed per unique trader across the four transaction categories: EOA–EOA, EOA–SC, SC–EOA, and SC–SC for Period 1. Similar plots for Periods 2 and 3 are shown in Figs. 2(a–d) and 3(a–d), respectively. Across all three periods, we analyze these four distinct categories of transaction types. In each figure, black markers represent raw data from individual trading accounts, while brown curves indicate log-binned averages. For log binning, 20 bins are created between the minimum and maximum number of unique trade partners, with bin size increasing on the log scale. Each data point is assigned to a bin, and the average values of trades within each bin are calculated to smooth out fluctuations. These brown curves highlight the overall scaling trend of the transactions. The scatter plots show a dense cluster of data points on the left side across all periods and transaction types. This implies that most of the trading accounts interact with only a few unique partners and carry out a small number of trades. Such a pattern reflects a heavy-tailed distribution typical of real-world networks, where most participants are low-activity users, and only a few accounts are highly active. The consistent presence of this pattern across all periods points to a stable core structure in trading behavior, despite variations in scaling across transaction types. Across all four categories and three periods, a positive correlation is observed. This suggests that accounts interacting with more unique partners tend to conduct more trades. However, the strength of this relationship, measured by the log-binned scaling exponent $\alpha$, varies by transaction type and period. The values of the exponent $\alpha$ obtained for different periods and transaction types are summarized in Table 2.

**Table 2.** Log-binned scaling exponents ($\alpha$) quantifying the relationship between the number of unique trade partners and trade volume for different transaction types across three time periods. A value of $\alpha \approx 1$ indicates near-linear scaling, suggesting that accounts with more partners conduct proportionally more trades. Lower $\alpha$ values, particularly for SC-SC transactions over time, reflect sublinear growth and increasing specialization or automation in contract-based interactions.

| Transaction Type | Period 1 | Period 2 | Period 3 |
|---|---|---|---|
| EOA–EOA | 1.01 | 0.98 | 1.00 |
| EOA–SC | 1.01 | 0.98 | 0.99 |
| SC–EOA | 0.98 | 0.95 | 0.94 |
| SC-SC | 0.93 | 0.78 | 0.67 |

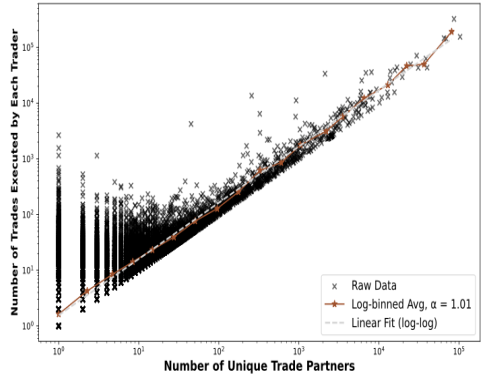

**(a)** EOA–EOA

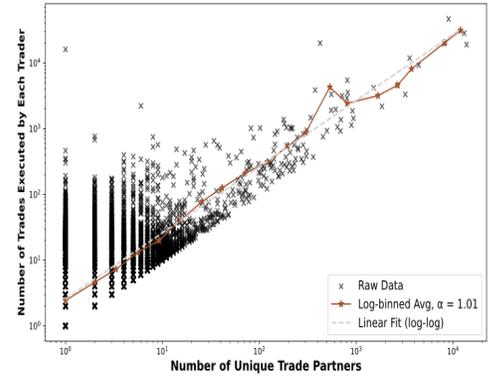

**(b)** EOA–SC

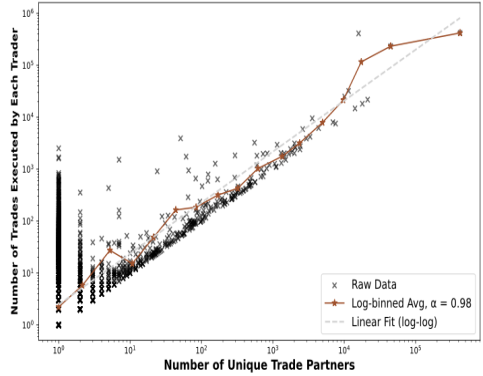

**(c)** SC–EOA

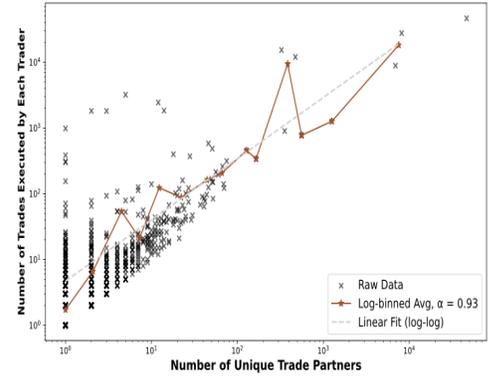

**(d)** SC-SC

**Fig 1.** In Period 1, Scatter plots in log-log scale depict the relationship between the number of trades executed per unique trader and the number of unique trade partners. Each subplot corresponds to one transaction type: (a) EOA–EOA, (b) EOA–SC, (c) SC–EOA, and (d) SC-SC. Black dots represent individual traders, while brown curves show log-binned averages, highlighting the underlying scaling trend. These visualizations provide insight into how trade activity scales against partner diversity.

From the scatter plot, we find that the EOA-related interactions EOA–EOA and EOA–SC show exponents close to unity ($\alpha \approx 1$) across all periods, indicating near-linear scaling. This near-linear scaling suggests that as EOAs connect with more partners, their activity increases proportionally. On the other hand, SC–EOA interactions show a slightly sublinear trend over time. The SC-SC interactions show a strong sublinear shift, with the scaling exponent decreasing from 0.93 in Period 1 to 0.67 in Period 3. This may indicate growing structural divergence in smart contract behavior, possibly due to increasing specialization or automation.

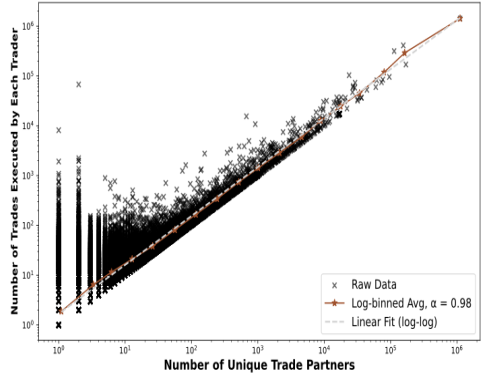

**(a)** EOA–EOA

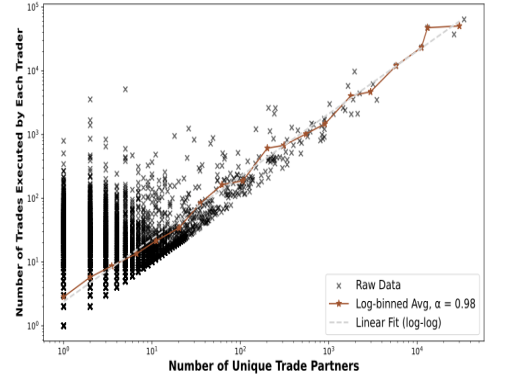

**(b)** EOA–SC

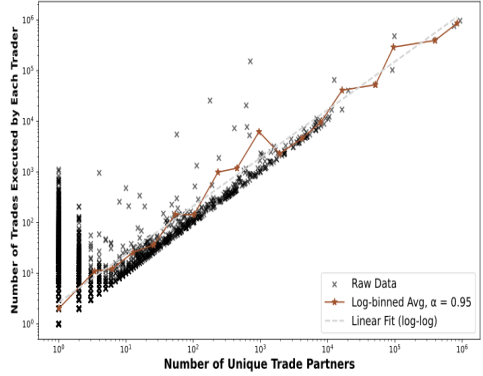

**(c)** SC–EOA

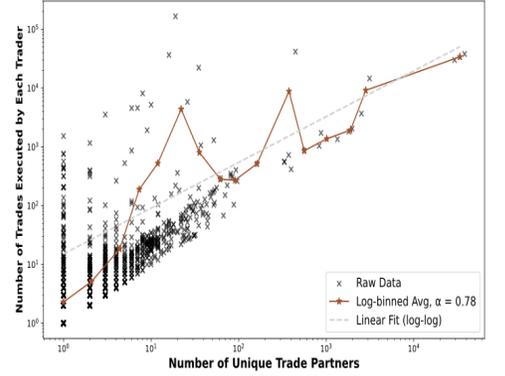

**(d)** SC-SC

**Fig 2.** Period 2: Scatter plots in log-log scale showing the relationship between the number of trades executed per unique trader and the number of unique trade partners across four categories during the middle period of our observation time window. Black markers indicate individual trading accounts, while the brown curves show log-binned averages that highlight the overall scaling behavior.

The consistent clustering of low-activity accounts, along with the observed scaling relationships, shows a clear heterogeneity in transaction behavior. EOA-related interactions, such as EOA–EOA and EOA–SC, show scaling exponents close to unity ($\alpha \approx 1$), indicating that as EOAs connect with more partners, their trading volume increases nearly proportionally. In contrast, SC–EOA interactions display a slightly sublinear trend, while SC-SC interactions show an increasingly sublinear pattern over time, with the exponent dropping from 0.93 in Period 1 to 0.62 in Period 3. This suggests that SC-SC behavior is becoming more structured or automated. These differences in scaling trends between EOA-based and SC-based transactions suggest that deeper, distinct behavioral patterns are present in the market. To further investigate whether these trends reflect universal properties, we further explore power-law fittings and TL outputs in the following subsection.

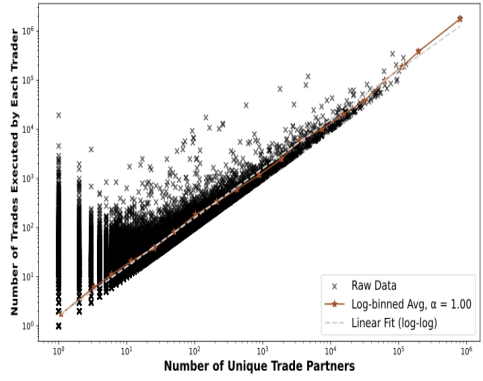
**(a)** EOA–EOA

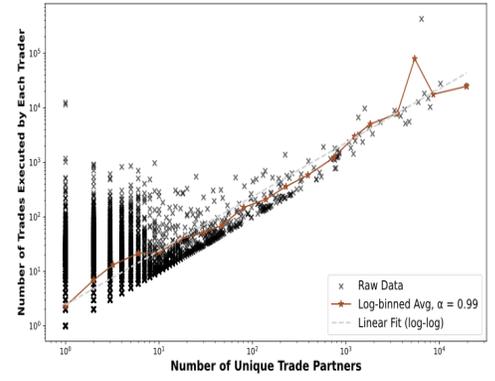
**(b)** EOA–SC

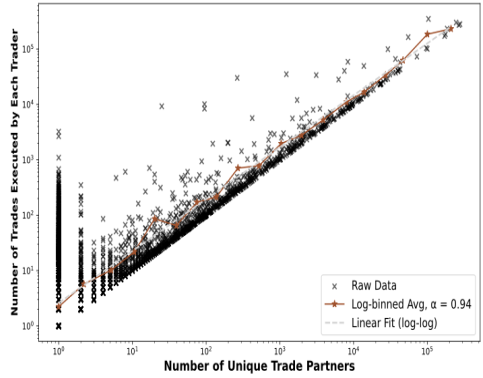
**(c)** SC–EOA

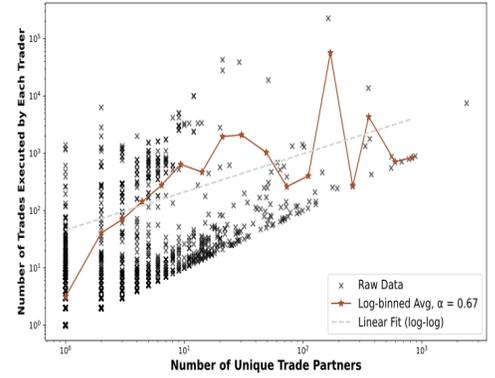
**(d)** SC–SC

**Fig 3.** Period 3: Scatter plots in log-log scale illustrating the relationship between the number of trades executed per unique trader and the number of unique trade partners across four categories. Black markers denote individual trading accounts, and the brown curves represent log-binned averages that reveal the overall scaling pattern. These plots help identify changes in interaction behavior in the final third of the observation window, particularly in SC dynamics.

### Power-law patterns in trading activity

To understand the trading behavior in ERC20 token transactions, we study whether the number of trades made or received by tokens follows a power-law distribution. We analyze four types of transactions: EOA-EOA, EOA-SC, SC-EOA, and SC-SC from both the sender and receiver sides for different periods. This helps us understand whether the activity patterns are similar for the sender and receiver. To study this, we apply a power-law model. In such a distribution, the probability of observing large values decreases slowly, often indicating the presence of a few highly active accounts. The scaling exponent $\gamma$ reflects the heaviness of the tail of the distribution. The lower values of $\gamma$ indicate the more extreme concentration, i.e., a few tokens dominate the trading activity. Table 3 summarizes the results of power-law fitting and values of the parameters for each transaction type over three distinct time periods considered in this study.

Figure 4 shows the log-binned empirical probability density and the corresponding power-law fits for EOA–EOA transactions across three different periods. Figures 4(a, b, and c) represent the sender distributions, while Figures 4(d, e, and f) show the receiver

**Table 3. Comparison of power-law fit results for sender and receiver across all transaction types and periods.**

| Transaction Type | Period | Role | $\gamma$ | $x_{\min}$ | Std. Error | KS Dist. | LLR | $p$-value |
|---|---|---|---|---|---|---|---|---|
| EOA–EOA | Period 1 | Sender | 2.32 | 5 | 0.0054 | 0.0188 | 21.08 | $1.11 \times 10^{-98}$ |
| | | Receiver | 2.63 | 10 | 0.0068 | 0.0121 | 12.53 | $5.09 \times 10^{-36}$ |
| | Period 2 | Sender | 2.26 | 4 | 0.0031 | 0.0202 | 31.20 | $9.85 \times 10^{-214}$ |
| | | Receiver | 2.55 | 10 | 0.0046 | 0.0051 | 16.37 | $2.94 \times 10^{-60}$ |
| | Period 3 | Sender | 1.76 | 105 | 0.0104 | 0.0127 | 9.11 | $8.54 \times 10^{-20}$ |
| | | Receiver | 2.53 | 13 | 0.0039 | 0.0097 | 25.37 | $5.47 \times 10^{-142}$ |
| EOA–SC | Period 1 | Sender | 2.44 | 6 | 0.0126 | 0.0171 | 7.31 | $2.71 \times 10^{-13}$ |
| | | Receiver | 2.07 | 11 | 0.0186 | 0.0145 | 3.89 | $9.85 \times 10^{-5}$ |
| | Period 2 | Sender | 2.48 | 14 | 0.0167 | 0.0116 | 5.40 | $6.70 \times 10^{-8}$ |
| | | Receiver | 2.00 | 18 | 0.0255 | 0.0178 | 2.53 | $1.14 \times 10^{-2}$ |
| | Period 3 | Sender | 2.47 | 9 | 0.0134 | 0.0115 | 8.07 | $6.90 \times 10^{-16}$ |
| | | Receiver | 1.89 | 23 | 0.0242 | 0.0143 | 7.14 | $9.37 \times 10^{-13}$ |
| SC–EOA | Period 1 | Sender | 1.92 | 12 | 0.0181 | 0.0124 | 4.29 | $1.79 \times 10^{-5}$ |
| | | Receiver | 1.92 | 3 | 0.0039 | 0.0358 | 14.20 | $8.90 \times 10^{-46}$ |
| | Period 2 | Sender | 1.75 | 19 | 0.0183 | 0.0213 | 8.34 | $7.70 \times 10^{-17}$ |
| | | Receiver | 2.52 | 7 | 0.0077 | 0.0095 | 15.49 | $3.92 \times 10^{-54}$ |
| | Period 3 | Sender | 1.61 | 19 | 0.0127 | 0.0362 | 16.76 | $5.11 \times 10^{-63}$ |
| | | Receiver | 2.43 | 3 | 0.0025 | 0.0139 | 23.52 | $2.47 \times 10^{-122}$ |
| SC-SC | Period 1 | Sender | 1.68 | 24 | 0.0604 | 0.0356 | 6.01 | $1.89 \times 10^{-9}$ |
| | | Receiver | 1.95 | 21 | 0.0461 | 0.0335 | 4.47 | $7.98 \times 10^{-6}$ |
| | Period 2 | Sender | 1.71 | 6 | 0.0268 | 0.0329 | 7.40 | $1.37 \times 10^{-13}$ |
| | | Receiver | 1.72 | 55 | 0.0423 | 0.0340 | 4.59 | $4.49 \times 10^{-6}$ |
| | Period 3 | Sender | 1.94 | 825 | 0.1148 | 0.0486 | 2.64 | $8.38 \times 10^{-3}$ |
| | | Receiver | 1.99 | 625 | 0.1056 | 0.0418 | 2.59 | $9.64 \times 10^{-3}$ |

Each row reports the scaling exponent ($\gamma$), minimum value threshold ($x_{\min}$) for power-law fitting, standard error (Std.Error), Kolmogorov–Smirnov (KS Dist.) distance, log-likelihood ratio (LLR) against an exponential distribution, and $p$-value for significance testing. Lower KS values and higher LLRs support in favour of a stronger fit to a power-law. A smaller $\gamma$ suggests a heavier tail, indicating that fewer traders dominate the transaction volume. These results highlight behavioral differences between EOAs and SCs, with contract-driven transactions showing more variation and extreme tail behavior across periods.

fits. As shown in Table 3, for Periods 1 and 2, the power-law exponents $\gamma$ for both sender and receiver distributions are above 2. The threshold parameter $x_{\min}$, which indicates where the power-law behavior begins, is low between 4 and 10. The standard errors are small, and the KS distances are also low. This suggests a close fit between the empirical distribution and the theoretical model. The LLR tests are all strongly positive and statistically significant. This confirms that the power-law model provides a significantly better fit than an exponential. Figure 5 shows the power-law distributions for EOA–SC transactions, where EOAs send tokens to SCs. A similar trend is observed for EOA-SC transactions with the EOA-EOA in Periods 1 and 2.

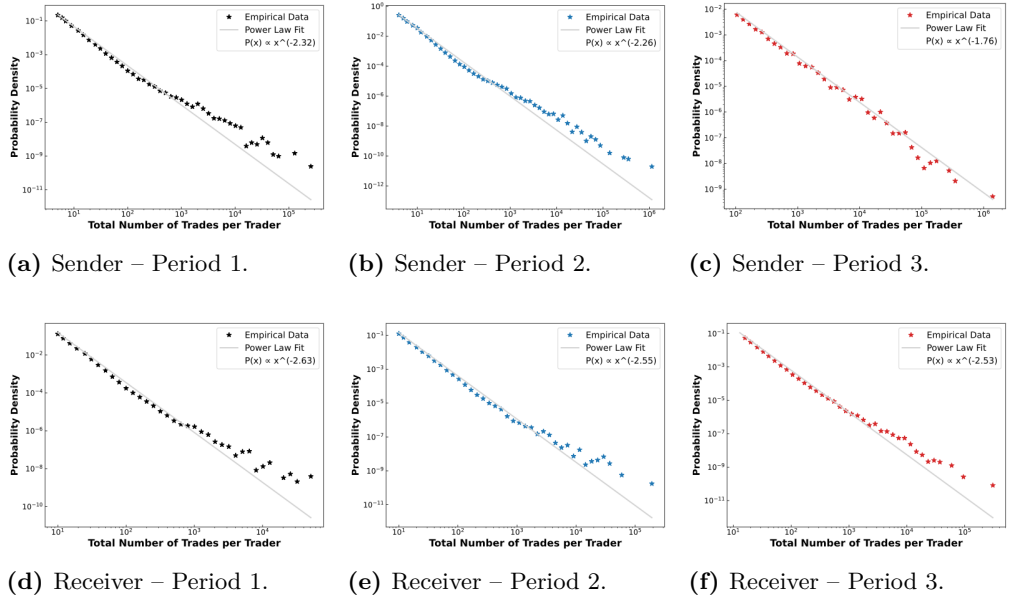

**(a)** Sender – Period 1. **(b)** Sender – Period 2. **(c)** Sender – Period 3.

**(d)** Receiver – Period 1. **(e)** Receiver – Period 2. **(f)** Receiver – Period 3.

**Fig 4.** Power-law fits of trading activities for **EOA–EOA** transactions across three time periods. The top row shows sender distributions and the bottom row shows receiver distributions, each on a log-log scale. Points represent log-binned empirical data, while the fitted lines illustrate the power-law distribution. These plots assess the heavy-tailed nature of trading frequency among EOAs, where a few accounts are responsible for a large number of trades. The slope of the fit $\gamma$ and the deviation in tails reveal behavioral consistency or variability across time.

During Period 3, the sender and receiver distributions for EOA–EOA transactions exhibit divergent behaviors. The sender distribution exhibits a lower exponent $\gamma = 1.76$ with a higher threshold $x_{\min} = 105$, while the receiver distribution remains consistent with earlier periods. A reverse behavior is seen in the EOA-SC transaction for the sender and receiver in Period 3 compared to EOA-EOA. The sender exponent is $\gamma = 2.47$ while the receiver has $\gamma = 1.89$. This asymmetric behavior during Period 3 may be linked with the crypto crash and postcrash conditions experienced in early 2018 [62]. This lower value of $\gamma$ during Period 3 for the sender in EOA–EOA and the receiver in EOA–SC also follows a power-law but with extreme concentration. These results indicate that a small number of accounts were responsible for a disproportionately large volume of transactions. One possible reason for this difference is that, after the crypto crash, individual EOAs become less active in sending tokens to each other, which leads to fewer but more dominant senders in EOA–EOA. Interaction with SCs also became more concentrated on a small set of popular contracts, resulting in highly skewed receiver activity in EOA–SC.

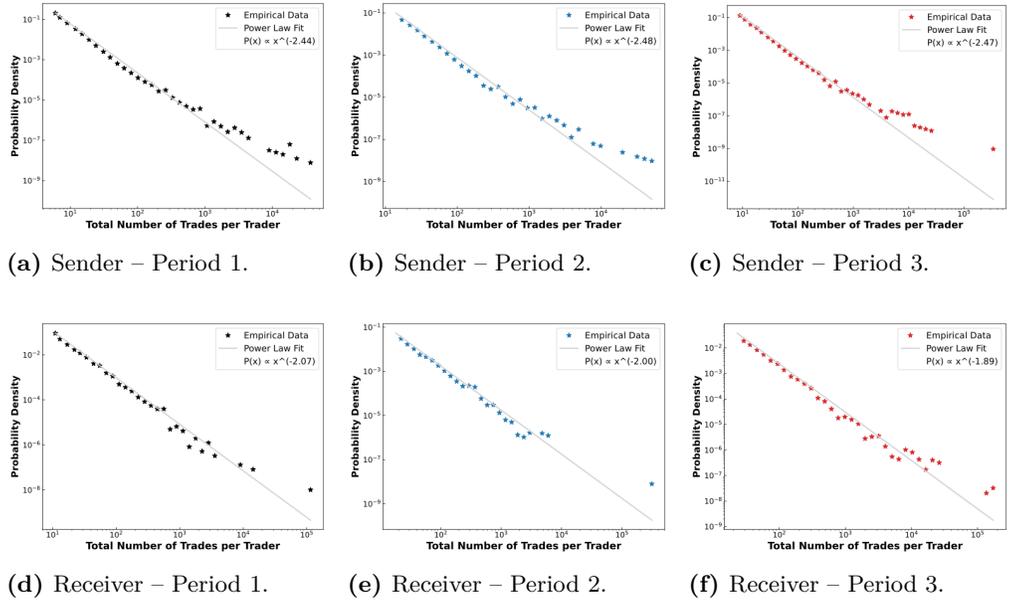

**(a)** Sender – Period 1.  **(b)** Sender – Period 2.  **(c)** Sender – Period 3.

**(d)** Receiver – Period 1.  **(e)** Receiver – Period 2.  **(f)** Receiver – Period 3.

**Fig 5.** Power-law fits of trading activity for **EOA–SC** transactions, where EOAs transfer tokens to SCs. The top row shows sender-side activity (EOAs) and the bottom row shows receiver-side activity (SC) across three time periods. Each subplot displays the empirical distribution and the corresponding power-law fit on a log-log scale. These plots help assess whether a small number of EOAs or contracts dominate trading volumes. Differences in fit quality, slope, and tail behavior across periods reflect shifting patterns in user interaction with SC platforms.

Figure 6 and 7 show the log-binned power-law fits for SC–EOA and SC-SC transactions across all three periods. For SC–EOA transactions, SC exhibits lower exponents $\gamma \in [1.61, 1.92]$, indicating heavy-tailed distributions with extreme concentration. While EOAs show variability, with $\gamma$ slightly below 2 in Period 1 and greater than 2 in Periods 2 and 3. This suggests a more balanced distribution of received transactions across EOAs. For SC–EOA, the lower $\gamma$ values for senders reflect that only a few SC were responsible for a large share of the outgoing transactions. This is likely due to automated or system-level behavior in specific protocols. SC-SC transactions not only show low $\gamma$ values ($\gamma \in [1.68, 1.99]$) for both sender and receiver, but also high $x_{\min}$ values. These $x_{\min}$ values are very high in Period 3, where thresholds reach 825 for senders and 625 for receivers. This indicates that only the most active SC, those engaging in frequent contract-to-contract interactions, contributed to the power-law behavior. Moreover, SC-SC interactions are characterized by higher standard errors and KS distances compared to other transaction types, suggesting that the empirical distributions deviate significantly from the theoretical power-law model. This could be due to the more complex, less predictable nature of contract-to-contract communication or the relatively smaller number of contracts actively participating in such interactions.

For all four transaction types: EOA–EOA, EOA–SC, SC–EOA, and SC-SC, the degree distributions for both senders and receivers consistently follow power-law behavior. This is evidenced by strongly positive and statistically significant LLR tests across all periods. This confirms that the power-law model offers a significantly better fit than an exponential. The values of the power-law exponent $\gamma$ and threshold $x_{\min}$ vary across different categories and periods. However, the presence of heavy-tailed distributions is a universal feature, reflecting the unequal and scale-free nature of transactional activity in the ERC20 ecosystem. Among all the transaction types, SC-SC transactions stand out

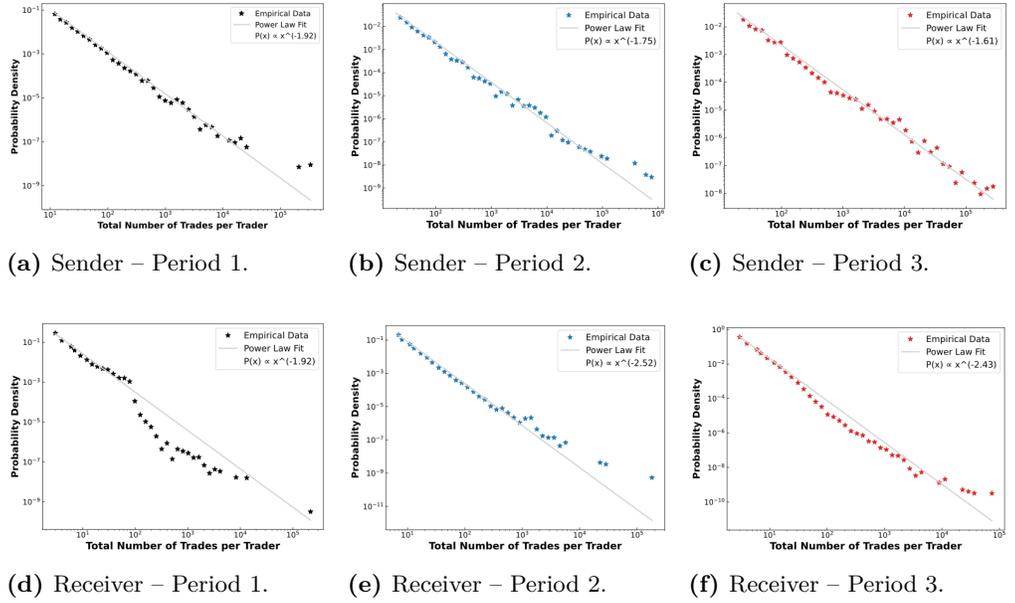

**(a)** Sender – Period 1.　**(b)** Sender – Period 2.　**(c)** Sender – Period 3.

**(d)** Receiver – Period 1.　**(e)** Receiver – Period 2.　**(f)** Receiver – Period 3.

**Fig 6.** Power-law fits of trading activity for **SC**–**EOA** transactions, where SCs send tokens to EOAs. The top row shows sender-side activity (SC), and the bottom row shows receiver-side activity (EOAs) across three distinct periods. Plotted on log-log scales, each subplot compares the empirical distribution of trades with a fitted power-law curve. These plots reveal how automated contract behaviors contribute to heavy-tailed activity patterns and how EOAs, as receivers, exhibit different levels of concentration and scaling over time.

with higher standard errors and KS distances, mainly in Period 3. This suggests a less stable fit and more variability in contract-to-contract interactions. Furthermore, their high $x_{\min}$ values indicate that only a small number of highly active contracts dominate these exchanges, highlighting a more centralized and system-driven interaction pattern. In contrast, transactions involving EOAs mainly on the receiver side tend to show higher $\gamma$ values and lower thresholds, indicating a broader and more scalable participation.

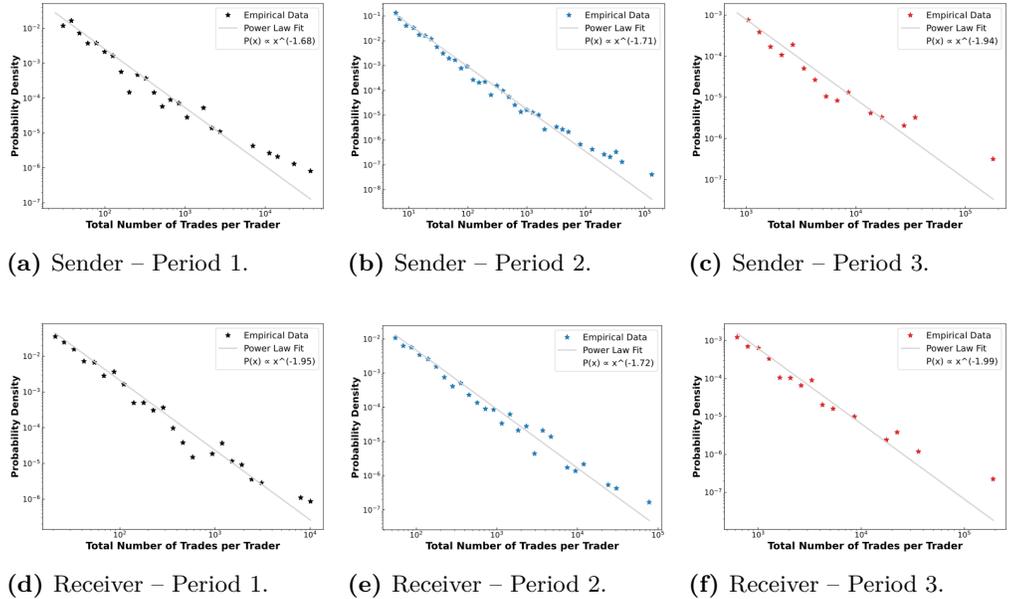

**(a)** Sender – Period 1. **(b)** Sender – Period 2. **(c)** Sender – Period 3.

**(d)** Receiver – Period 1. **(e)** Receiver – Period 2. **(f)** Receiver – Period 3.

**Fig 7.** Power-law fits of trading activity for **SC-SC** transactions, where both sender and receiver are SC. The top row shows sender-side contract activity, and the bottom row shows receiver-side contract activity over three time periods. Plotted on log-log scales, these subfigures display the empirical distributions of trade frequencies alongside their power-law fits. These plots highlight the behavior of highly active, automated contract-to-contract interactions, often marked by heavy tails, high thresholds, and variability across time. Such patterns reflect systemic processes or protocol-level dynamics in SC ecosystems.

These power-law findings support the earlier results from the analysis of the scaling relationship between trade volume and partner diversity in Subsection 1.1. Both analyses indicate that user-based EOA-driven transactions exhibit more universal and scalable behavior, while contract-driven activities, particularly SC-SC, appear more context-dependent and structurally variable. The consistency between these two complementary approaches strengthens our confidence in the observed behavioral distinctions between users and automated entities in the ERC20 network. To check whether these differences follow a common pattern, we examine the stationarity of the time series data, followed by an application of Taylor's law.

## Stationarity Test Results

Before applying TL to examine how the variability in trading activity scales with their mean. Prior to TL analysis, stationarity in the time series data is ensured by the KPSS test. We have chosen a 1-hour time window to derive the temporal TL exponent in all ERC20 transaction types, EOA–EOA, EOA–SC, SC–EOA, and SC–SC. For each trader, we constructed a time series representing their trading activity per hour. The KPSS test was applied individually to each of these time series. A p-value greater than 0.05 indicates that the time series is stationary, meaning we do not reject the null hypothesis. The table below summarizes the percentage of traders in each transaction category whose time series met the stationarity condition based on their p-values. As shown in Table 4, for all periods and transaction types, over 90% of traders showed stationary behavior. This confirms that the time series data suits TL.

**Table 4.** Percentage of traders exhibiting stationary trading behavior based on the KPSS test, applied to hourly trading activity across three time periods. A result above 90% indicates that the majority of traders in each transaction category maintained stable behavior over time. These results validate the use of Taylor's law by confirming the stationarity assumption for time series analysis in most cases.

| Transaction Type | Period 1 | Period 2 | Period 3 |
|---|---|---|---|
| EOA-EOA | 99% | 98% | 99% |
| EOA-SC | 92% | 94% | 95% |
| SC-EOA | 99% | 97% | 100% |
| SC-SC | 97% | 98% | 93% |

## Temporal TL Exponent

We employed temporal TL to further evaluate whether the trading behavior observed in ERC20 token interactions reflects universal patterns. This method examines whether there is a consistent relationship between the mean number of trades made by an account and the degree of fluctuation in its activity over time. Table 5 shows the value of the exponents ($\beta$) of Taylor's law for different situations. The exponent $\beta$ quantifies the strength of the scaling relationship. It captures the relationship between the temporal mean and variance of token trades for individual tokens. We calculate the value of $\beta$ for both the sender and receiver and each transaction type: EOA–EOA, EOA–SC, SC–EOA, and SC-SC across the three time periods.

**Table 5.** TL exponents $\beta$ measure the scaling relationship between variance and mean of hourly trading activity across different transaction types and roles over three time periods. The slope $\beta \approx 2$ suggests near-quadratic scaling, indicating that variability in trading activity grows proportionally with the mean. This table helps identify consistency, divergence, and structural differences in user- and contract-driven behaviors.

| Transaction Type | Role | Period 1 ($\beta$) | Period 2 ($\beta$) | Period 3 ($\beta$) |
|---|---|---|---|---|
| EOA–EOA | Sender | 2.35 | 2.36 | 2.35 |
|  | Receiver | 2.31 | 2.31 | 2.19 |
| EOA–SC | Sender | 2.28 | 2.37 | 2.19 |
|  | Receiver | 2.26 | 2.26 | 2.36 |
| SC–EOA | Sender | 2.42 | 2.51 | 2.39 |
|  | Receiver | 2.31 | 2.19 | 2.14 |
| SC–SC | Sender | 2.15 | 1.96 | 2.09 |
|  | Receiver | 2.47 | 2.27 | 2.24 |

For all categories and periods, the values of $\beta \in [1.96, 2.51]$. Figures 8, 9, 10, and 11 show the plot between the log mean *vs* log variance for EOA–EOA, EOA–SC, SC–EOA, and SC-SC transactions, respectively. In the EOA–EOA transaction, the $\beta$ values are consistent for all periods. Both the sender and receiver $\beta$ values remain stable with $\beta_{\text{sender}} = 2.35 \pm 0.01$ and $\beta_{\text{receiver}} = 2.27 \pm 0.06$. This near-quadratic scaling ($\beta \approx 2$) suggests a predictable relationship, where doubling the mean activity results in a fourfold increase in variance. These results suggest that individuals using EOAs maintain regular patterns of trading over time, with increasing trade frequency showing proportional variability. Similarly, EOA–SC interactions show stable $\beta$ in Periods 1 and 2. In Period 3, however, the sender exponent decreases to $\beta_{\text{sender}} = 2.19$ whereas the

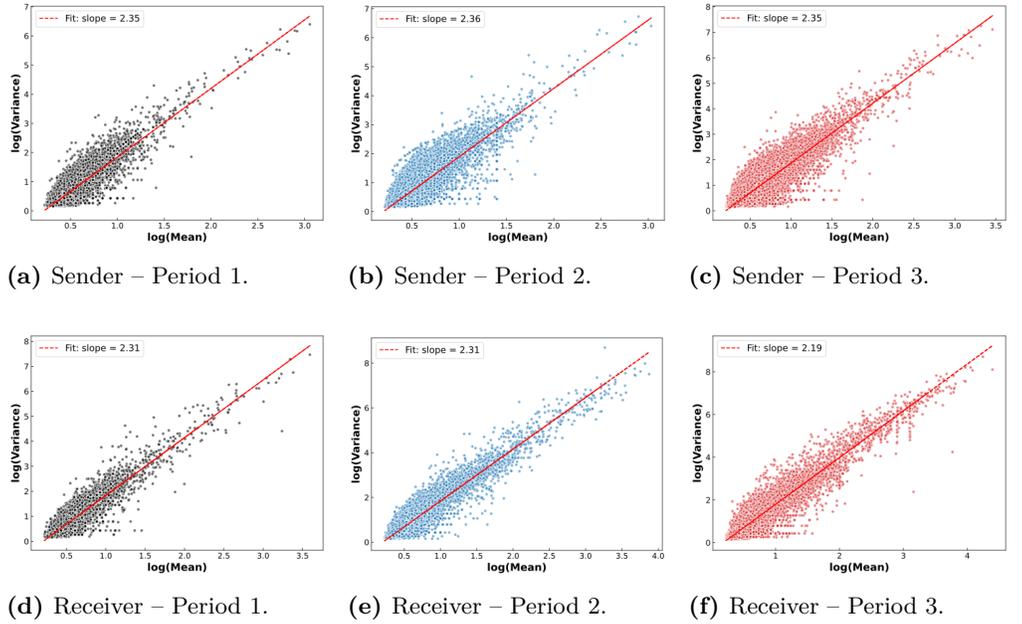

**(a)** Sender – Period 1.    **(b)** Sender – Period 2.    **(c)** Sender – Period 3.

**(d)** Receiver – Period 1.    **(e)** Receiver – Period 2.    **(f)** Receiver – Period 3.

**Fig 8.** Scatter plot of log(mean) versus log(variance) for **EOA–EOA** transactions. The top and bottom rows illustrate, respectively, sender and receiver activities during three observation time windows. The slope of the linear fit represents the temporal TL exponent $\beta$, which captures how trading variability scales with average activity. Consistent $\beta$ values ($\sim 2$) for senders across three periods, suggesting predictable user-driven trading behavior over time.

receiver exponent increases to $\beta_{\text{receiver}} = 2.36$. This likely reflects that, after the crypto market crash, token deposits became concentrated in a small number of heavily used contracts.

From table 5, we see that SC-EOA and SC-SC, $\beta$ show divergent trends. In the SC–EOA, sender exponent increases from $\beta_{\text{sender}} = 2.42$ in Periods 1 to 2.51 in Period 2, whereas the receiver exponent contracts reduce to $\beta_{\text{receiver}} = 2.19$. This pattern suggests that SC increasingly dominated outgoing flows via automated protocols, while recipient EOAs diversified their activity profiles. SC-SC interactions show the most variation of $\beta$ values over time. During Period 2, the sender exponent drops to $\beta_{\text{sender}} = 1.96$, suggesting that the usual relationship between average activity and its variability breaks down temporarily; however, by Period 3, it recovered to 2.09. On the other hand, the receiver exponents remain relatively high, ranging from 2.24 to 2.47, reflecting consistent protocol-driven inflows.

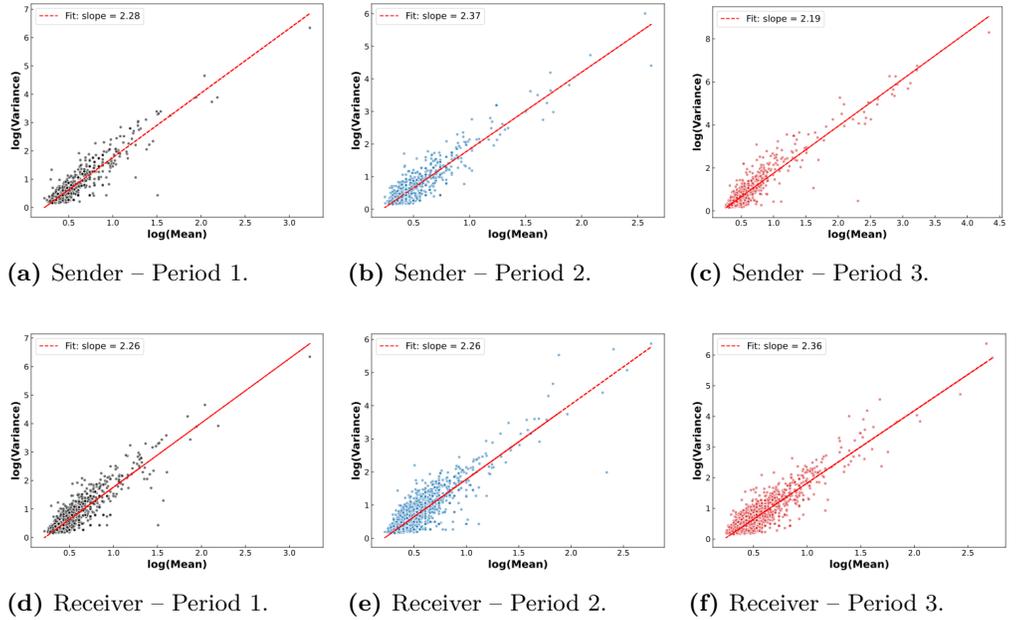

**(a)** Sender – Period 1.  **(b)** Sender – Period 2.  **(c)** Sender – Period 3.

**(d)** Receiver – Period 1.  **(e)** Receiver – Period 2.  **(f)** Receiver – Period 3.

**Fig 9.** Temporal TL exponents, derived for **EOA–SC** transactions, are shown across three periods. The top and bottom rows show sender activity (EOAs sending tokens to SC) and receiver activity (SC receiving tokens), respectively. Each point represents a trader. The linear fit of the scatter plot between log-mean versus log-variance of hourly trade activity represents TTL exponent $\beta$. Consistent $\beta$ ($\sim 2$) values for senders across three periods, suggesting predictable user-driven trading behavior over time.

To better understand how user behavior influences the observed TL exponents, we compare the scaling relationships in Table 2 with the power-law exponents in Table 3. For EOA-driven transactions, $\beta$ remains close to 2, consistent with near-linear scaling ($\alpha \approx 1$) between trade partners and volume in Table 2. This suggests balanced growth in user activity as the number of partners and trade volume increase together. SC-SC interactions exhibit lower $\alpha$ and $\gamma$ values, reflecting sublinear scaling and a shift toward automation and centralization. These comparisons across tables reveal that shifts in partner diversity and heavy-tailed trading behavior play a key role in shaping the observed scaling patterns.

EOA roles are tightly grouped around $\beta = 2.3$ for senders and $\beta = 2.2$ for receivers. This indicates that interactions driven by human users tend to follow consistent and near-universal statistical patterns. Smart contract roles, on the other hand, show much variation. The range of $\beta$ values for SC-SC interactions is $\Delta\beta_{\text{Smart--Smart}} = 0.51$, compared to just $\Delta\beta_{\text{EOA--EOA}} = 0.17$ for EOA–EOA interactions. This suggests that smart contract behavior is more influenced by the specific context or logic in which they operate. These findings support our observations on scaling behavior. Interactions driven by people are more predictable and stable, while those managed by algorithms depend heavily on external factors like protocol updates, market conditions, and the growth of decentralized finance(DeFi). Together, these trends highlight how both human-driven and algorithm-driven forces shape activity in the ERC20 ecosystem.

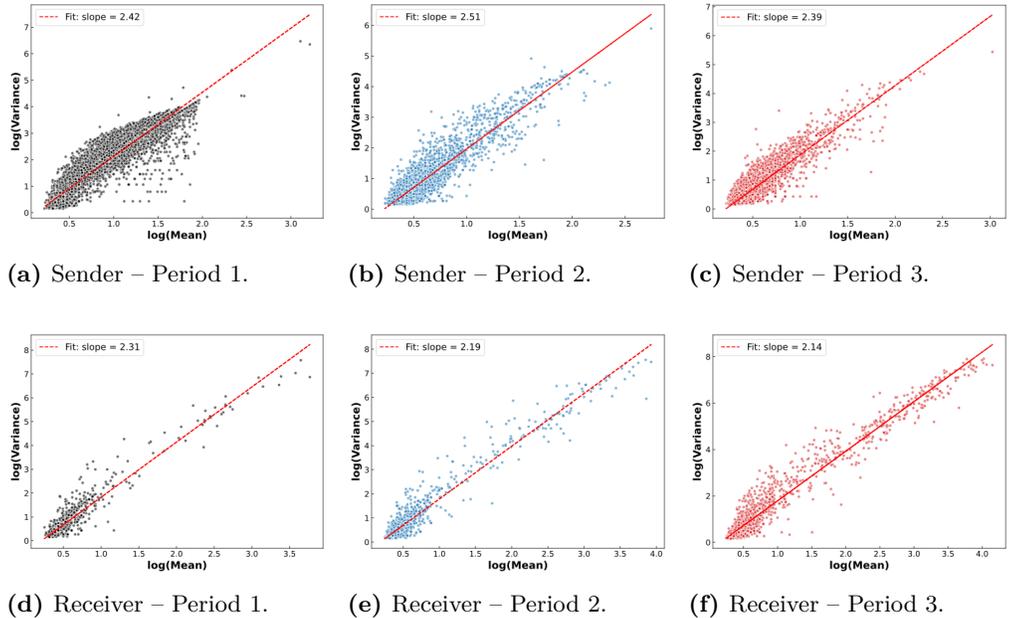

**Fig 10.** Temporal TL for **SC**–**EOA** transactions. The top row shows sender activity (SC sending tokens to EOAs), and the bottom row shows receiver activity (EOAs receiving tokens), across three periods. Each point corresponds to a trader, with log-mean and log-variance plotted. The slope of the fitted line represents the Taylor exponent $\beta$, indicating how variability in activity scales with the average. These plots highlight the dynamics of automated token distributions and user reception patterns over time.

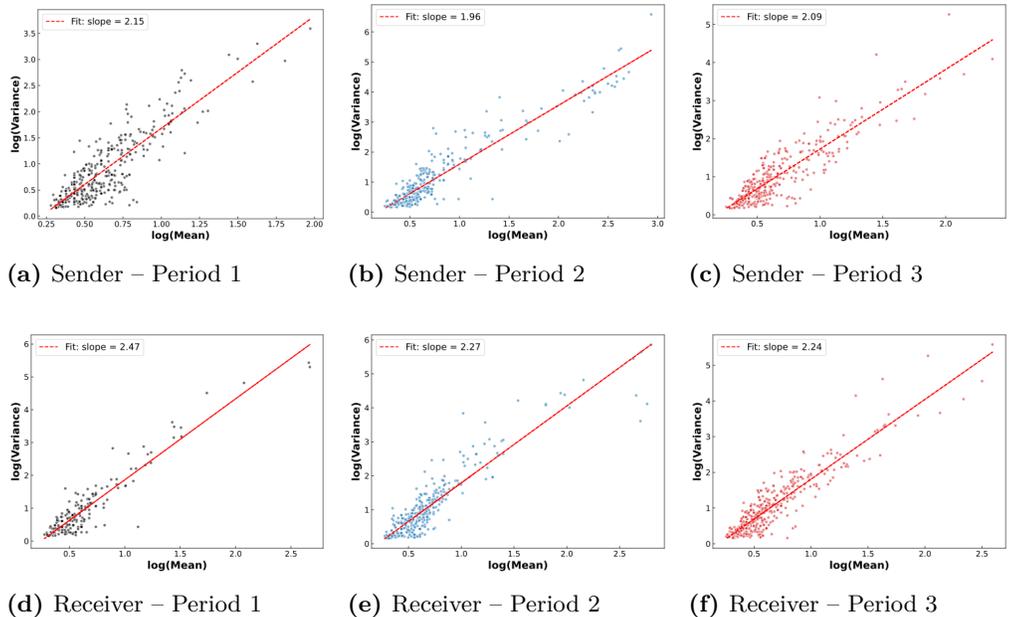

**Fig 11.** Temporal TL for **SC**–**SC** transactions. The top row shows sender-side activity (SC initiating transfers), while the bottom row shows receiver-side activity (SC receiving tokens) across three time periods. Each point plots the log-mean versus log-variance of hourly trade volume for each contract. The slope of the line is the Taylor exponent $\beta$, which reveals how volatility scales with mean activity. Compared to other transaction types, these plots show greater variation, indicating more diverse and protocol-specific behaviors among SC.

On the other hand, SC-SC, and SC–EOA interactions display notable deviations from the theoretical line, especially in the upper quantiles during Periods 2 and 3. These departures indicate the presence of heavy-tailed or skewed residual distributions, suggesting that the Taylor scaling model captures only part of the underlying structure. This aligns with the wider range of $\beta$ values observed for SC roles, such as $\Delta\beta_{\text{Smart–Smart}} = 0.51$, and reflects the diverse, protocol-specific dynamics of contract-based activity. Smart contracts, driven by automated mechanisms, exhibit bursty or concentrated flows that deviate from the more uniform patterns typical of human user behavior.

This clear difference supports our findings, as EOA interactions follow universal patterns, while SC behavior is context-dependent. Human activity is predictable, while algorithmic interactions depend on external factors like protocol rules or market conditions.

## Conclusion

We analyzed transactional behaviors of ERC20 tokens on the Ethereum blockchain by categorizing interactions into four typesEOA–EOA, EOA–SC, SC–EOA, and SC-SC based on whether human-controlled EOAs or algorithmically governed SCs initiated them. The central goal was determining whether these interaction types exhibit consistent and universal statistical signatures over time. Our results show that EOA-driven transactions display highly regular, scalable, and predictable behavior, as evidenced by near-linear scaling relationships, robust power-law distributions with heavy tails, and stable TL exponents across all periods analyzed. These findings indicate that user-driven blockchain transactions follow universal patterns observed in traditional economic and social systems. In contrast, SC-driven interactions, particularly SC–SC exchanges, exhibited greater variability, characterized by sublinear scaling, unstable and threshold-sensitive power-law exponents, and frequent deviations from theoretical models. These irregularities were further reflected in elevated KS distances and larger standard errors. These patterns suggest that smart contract behavior is not governed by universal dynamics but is instead shaped by contextual factors such as protocol design, automation logic, and transaction purpose.

Overall, our findings uncover statistically distinctive features of human- versus algorithm-driven interactions on the ERC20 network. While user-driven transactions align with universal behavioral laws observed in traditional economic and social systems, automated SC interactions tend to be more context-dependent and structurally volatile. This distinction highlights the importance of adopting separate analytical frameworks for modeling human and algorithmic behavior in decentralized systems. These insights are not only critical for designing resilient DeFi infrastructures but also offer valuable indicators for monitoring systemic risk and volatility in emerging crypto-financial markets.

In future work, we plan to extend this framework across multiple blockchain platforms (e.g., Binance Smart Chain, Solana, Avalanche) to evaluate whether similar universality patterns persist beyond Ethereum. Moreover, exploring causal mechanisms behind smart contract volatility, such as governance rules, transaction fee dynamics, or oracle dependencies, can offer deeper insights into protocol-level fragilities. Lastly, as a future scope of research, we plan to integrate machine learning techniques to classify behavioral regimes and anomalies at scale, opening the path toward intelligent monitoring of DeFi ecosystems.

## Acknowledgments

The authors acknowledge the National Institute of Technology Sikkim for allocating doctoral research fellowships to Kundan Mukhia and S.R. Luwang.